\documentclass[journal=jacsat,manuscript=article]{achemso}

\usepackage[version=3]{mhchem} 

\author{Hans Kleemann}
\email{Hans.Kleemann@iapp.de}
\affiliation{Institut f\"ur Angewandte Photophysik, Technische Universit\"at Dresden, 01069 Dresden, Germany}
\author{Rafael Gutierrez}
\affiliation{Institute for Materials Science and Max Bergmann Center for Biomaterials, Technische Universit\"at Dresden, 01069 Dresden, Germany}
\author{Frank Lindner}
\affiliation{Institut f\"ur Angewandte Photophysik, Technische Universit\"at Dresden, 01069 Dresden, Germany}
\author{Stanislav Avdoshenko}
\affiliation{Institute for Materials Science and Max Bergmann Center for Biomaterials, Technische Universit\"at Dresden, 01069 Dresden, Germany}
\author{Pedro D. Manrique}
\affiliation{Institute for Materials Science and Max Bergmann Center for Biomaterials, Technische Universit\"at Dresden, 01069 Dresden, Germany}
\author{Bj\"orn L\"ussem}
\author{Gianaurelio Cuniberti}
\affiliation{Institute for Materials Science and Max Bergmann Center for Biomaterials, Technische Universit\"at Dresden, 01069 Dresden, Germany}
\alsoaffiliation
{National Center for Nanomaterials Technology, POSTECH, Pohang 790-784, Republic of Korea}
\author{Karl Leo}
\affiliation{Institut f\"ur Angewandte Photophysik, Technische Universit\"at Dresden, 01069 Dresden, Germany}

\title{Organic Zener Diodes: Tunneling across the Gap in Organic Semiconductor Materials.}
\begin{document}
\begin{abstract}
Organic Zener diodes with a precisely adjustable reverse breakdown from -3 V to -15 V without any influence 
on the forward current-voltage curve are realized. This is accomplished by controlling the width of the charge depletion zone in a $pin$-diode with an accuracy of one nanometer independently of the doping concentration and the thickness of the intrinsic layer. The breakdown effect with its exponential current voltage behavior and a weak temperature dependence is explained by a tunneling mechanism across the HOMO-LUMO gap of neighboring molecules.
The experimental data are confirmed by a minimal Hamiltonian model approach, including coherent tunneling 
and incoherent hopping processes as possible charge  transport pathways through the effective device region. 
\end{abstract}

In the last few years, the physical properties of organic semiconductors have been studied with considerable detail.
Despite the fact that there is still a limited understanding of the relevant microscopic charge transport scenarios in such systems, 
various organic electronic devices such as efficient organic
light-emitting diodes (OLEDs) {\cite{Tang1987,Burroughes1990}}, 
organic memory cells {\cite{Ouyang2004}}, organic 
field-effect transistors {\cite{Horowitz1998}}, vertical triodes 
{\cite{Yang1994}}, and high-frequency diodes {\cite{Steudel2005}} 
have been successfully demonstrated. 

The main difference between inorganic and organic semiconductors is that the latter
usually do not display (at room temperature) band-like transport through delocalized states, but  rather exhibit 
hopping of carriers between spatially localized states. This can either happen as electron transport
through the lowest unoccupied molecular orbitals (LUMO) or hole transport through the highest occupied orbitals
(HOMO). Typically, this results in mobilities three or more orders of magnitude below those of band transport 
and an entirely different temperature dependence, i.e. an increase of mobility with increasing temperature can be obtained.

In inorganic semiconductors, there are a number of basic transport effects which involve more than one band, e.g. 
interband (Zener) tunneling or avalanche breakthrough.~\cite{Sze1966,McAfee1951}
While these effects are now understood in detail for inorganic semiconductors, they have not been experimentally studied 
in detail for organic semiconductors so far. Due to the structural complexity of organic semiconductors, 
the theoretical study of their charge transport properties is also quite challenging, so that 
theoretical studies are usually limited to structurally ordered systems.~\cite{troisi2010,PhysRevLett.102.116602,citeulike:2842232,PhysRevB.79.235206}
A key electronic device based on the Zener effect, the Zener diode, is crucial for basic electrical 
circuit requirements, such as voltage- and temperature stabilization and over-voltage protection. An organic Zener diode with 
an adjustable breakdown is still missing. Likewise Zener diodes are key elements for passive matrix memory arrays, {\cite{Scott2007}}
where Zener diodes are essential to prevent parasitic current flow through non-selected crosspoints. The required parameters 
for the Zener diode are given by the read-, write- and erase voltages of the memory. {\cite{Lindner2008}} 

In the Letter, we investigate the HOMO-LUMO Zener tunneling in an organic semiconductor system by detailed 
measurements of the reverse currents in organic $pin$-diodes. In particular, we provide evidence for  the tunneling nature
of the reverse currents by showing a strong influence of the effective tunnel barrier and a very weak  temperature dependence.
To get a first insight into the possible charge transport mechanisms, we complement the experimental investigations by  an effective low-dimensional model which is then treated in two limiting physical situations: incoherent hopping transport and coherent tunneling. Our results suggest that both mechanisms may exist in the system under study in dependence of the sign of the applied voltage.


\textit{Experimental part.~} We address  Zener tunneling in the reverse current regime of organic $pin$ diodes.
The breakthrough properties of such devices have only been rarely investigated, see e.g.~\cite{Liang2003}. 
The advantage of the $pin$-diode concept~{\cite{Harada2005}}
 is that the geometry and the potentials within the semiconductor layers can be controlled 
very precisely, thus allowing to study the tunneling
effects, which sensitively depend on minute variations of these parameters, in a very controlled manner.
Thus, the tunneling current is independently tunable by two parameters: the thickness 
of the intrinsic interlayer (IIL), and the doping concentration of the electron and hole transport 
layers (ETL/HTL). Doping changes the charge carrier concentration~{\cite{Walzer2007}}
 and width of depletion regions.~{\cite{Blochwitz2001}} Furthermore, one obtains a barrier-free transport until the charge carriers reach the IIL.~{\cite{Olthof2009}} In contrast, the IIL acts as an additional 
depletion layer and the voltage drop over the $pin$-junction 
mainly occurs in this intrinsic layer.~{\cite{Olthof2009}} A comparative study of organic $pn$- and $pin$-junctions comprising doped organic layers would be desirable. However, as we will show in this letter the typical width of the native
depletion zone in the doped layers is less than 3 nm. Therefore, the current in a $pn$-junction would be caused not only by drift and diffusion but also by a tunneling mechanism. In this way the backward current should not be tunable independently of the forward direction.   

\begin{figure}
\includegraphics[scale=0.6]{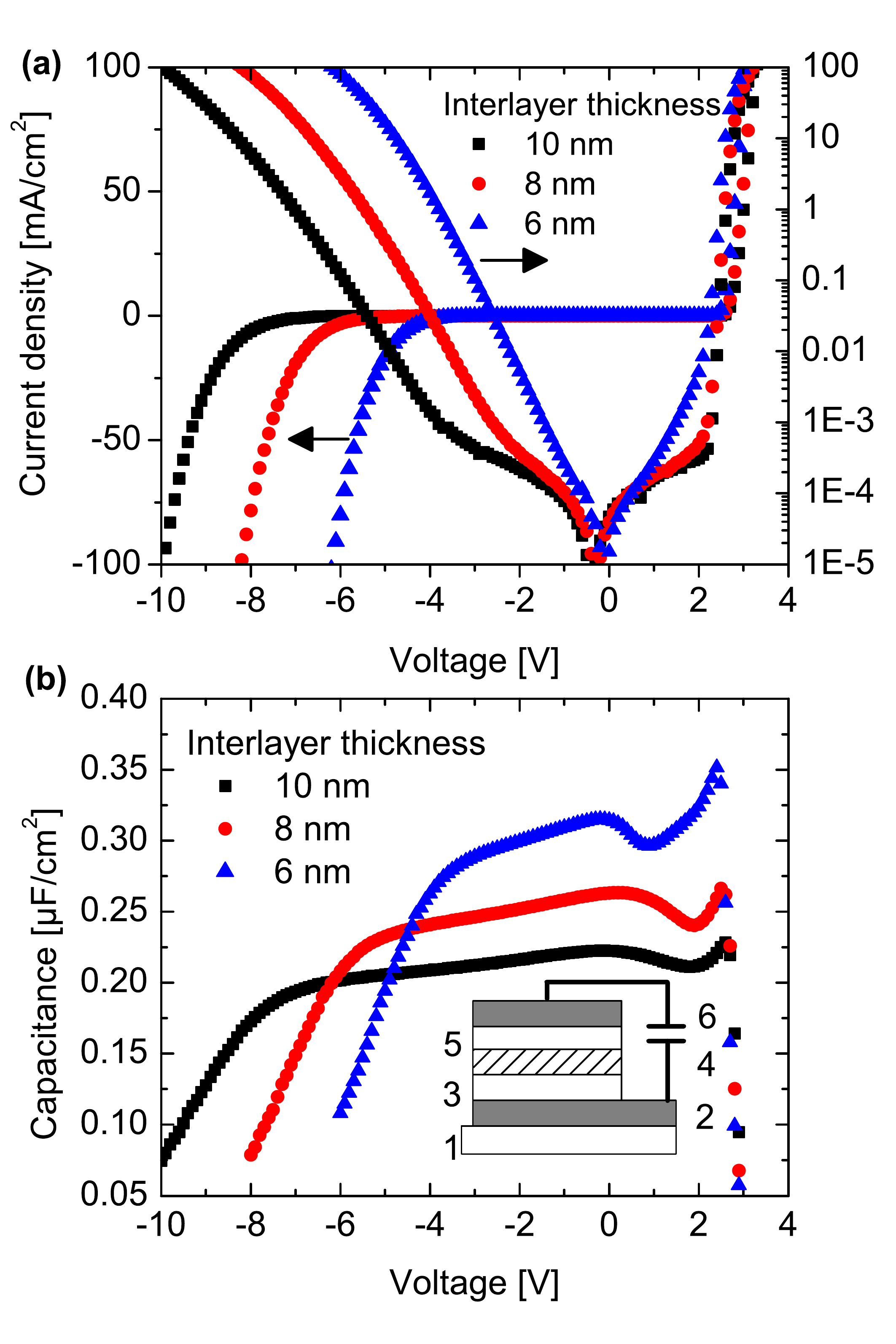}
\caption{ Current voltage characteristics
for different interlayer thicknesses (b) Capacitance 
voltage characteristics (measured  at 1 kHz) of the samples 
shown in \ref{fig_1}a. The inset is a sketch of the device 
structure and the layers are listed as they are evaporated 
on a cleaned glass substrate (1):
Al(100nm)(2) /MeO-TPD:F4TCNQ[4wt\%](50nm)(3) 
/Balq2:NPB[50wt\%](4) /Bphen:Cs(50nm)(5) /Al(100nm)(6)}
\label{fig_1}
\end{figure}

The structures investigated (see the inset of \ref{fig_1}b)  consist of 
three organic layers sandwiched between two aluminium 
contacts, which define the active area of 6.38 mm$^2$. 
The inner structure of the organic layers follows the 
$pin$-concept with a highly p-type doped HTL, a highly n-type doped ETL, 
and an intrinsic layer in between (see Supporting Information for details).  
\ref{fig_1} displays the current-voltage ($I$-$V$) characteristics 
of such a device. In forward direction the expected 
diode-like behavior~{\cite{Harada2005}} 
is observed. The interlayer thickness does not affect the onset and 
the slope of the current rise in forward direction, which 
appears at 2.3 V. In contrast, there is a significant 
influence on the reverse current, which can be seen in 
the loss of the rectifying property at a certain voltage 
depending on the interlayer thickness. The onset of the 
reverse breakdown shifts in a controlled manner 
to higher negative voltages (from -3 V to -15 V) by 
increasing the interlayer thickness (see \ref{fig_1}a): we 
observe a shift of the reverse breakdown voltage of 
approximately 1 V by each additional nanometer of 
 IIL thickness. Current densities of more than 
300 mA/cm$^2$ in reverse direction can be achieved without 
limiting reversibility or reproducibility. Obviously, the large reverse voltage region is dominated 
by a purely exponential rise of the current, which provides a strong hint for a tunneling like mechanism. 
Also a flat part of the $I$-$V$ curve can be observed in reverse direction. Since this is more pronounced for thicker interlayers it is considered
to be related to a leakage currents. Hence, a larger field and a larger voltage are needed to be in the tunneling regime for thicker interlayers.  
The limitation of the transport layers and the injection gives rise to the flattening of the $I$-$V$ curve for larger negative voltages. 

\begin{figure}
\includegraphics[scale=0.6]{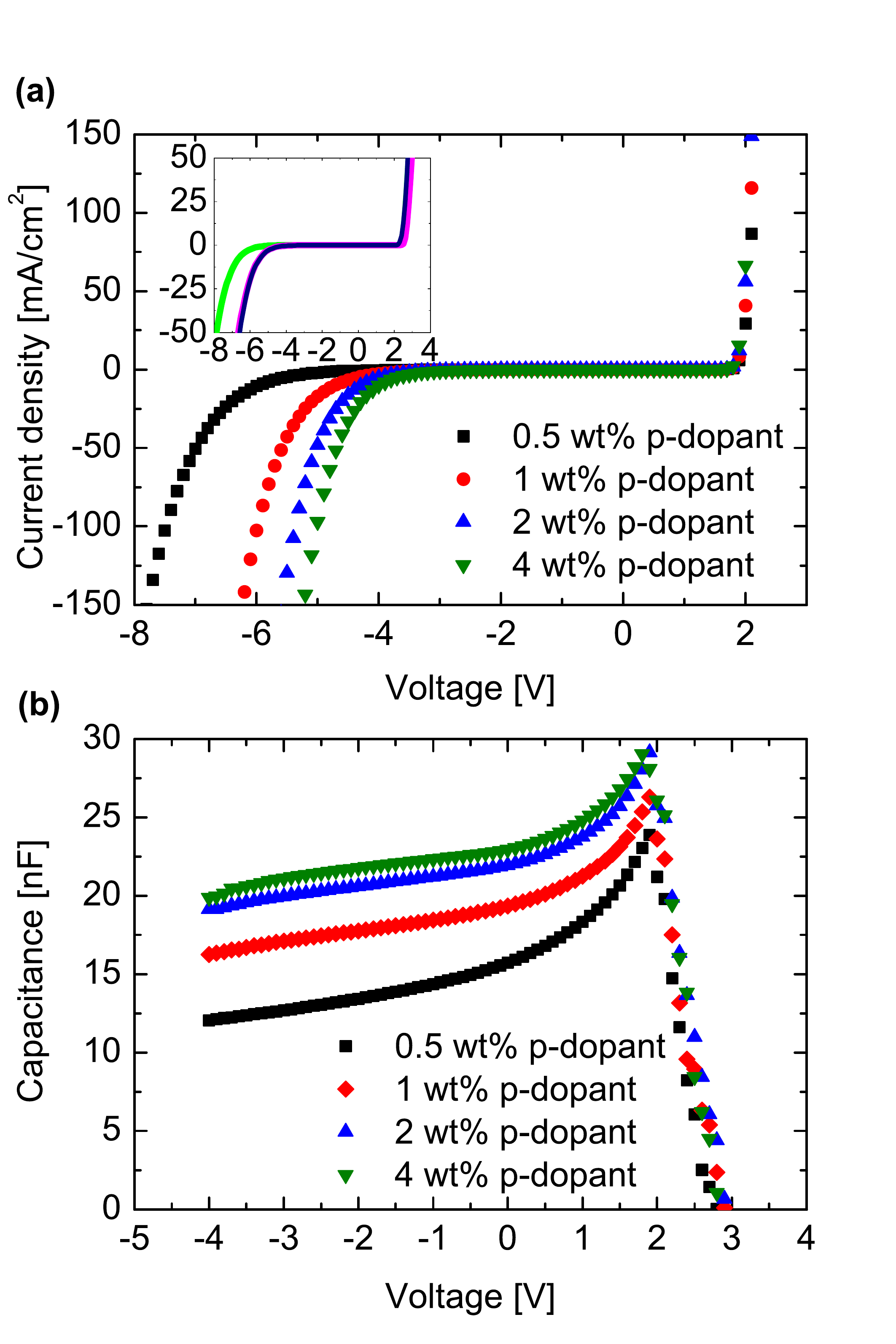}
\caption{ (a) Current density for different doping 
concentrations in the HTL. The devices consist 
of Al(100nm)/ RE68:NDP2[x wt\%](50nm)/ RE68(7nm)/ 
RE68:NDN1[16wt\%](50nm)/ Al(100 nm) on a cleaned 
glass substrate. Inset: comparison of different dopand 
materials regarding the $I$-$V$ performance. The structure consists 
of a doped HTL(50nm), an intrinsic interlayer (Balq2:NPB[50wt\%]
(7nm)) and a doped ETL(50nm). The colors denote (HTL/ETL): 
Meo-TPD:F4TCNQ[4wt\%]/ BPhen:Cs (light green), 
Meo-TPD:NDP2[4wt\%]/ BPhen:Cs (magenta), Meo-TPD:NDP2[4wt\%]/ 
BPhen:NDN1[16wt\%] (dark blue) (b) Capacitance voltage 
characteristics of the samples shown in \ref{fig_3}a}\label{fig_3}
\end{figure}

To further understand these observations, we perform capacitance-voltage ($C$-$V$) measurements, 
allowing to study the voltage-dependent effects of 
charge depletion and accumulation in the IIL and allow 
correlating layer thicknesses to capacitance values. 
In our devices, the transport layers have a larger thickness 
and a better conductivity in comparison to the IIL which is 
thus the dominating $RC$-element in the stack. Capacitance 
frequency ($C$-$f$) measurements are done in order to prove 
independently the capacitance calculated by the 
assumption of one dominant $RC$-circuit (see Supporting Information). Thus, it is checked 
that the calculated capacitance at 0 V corresponds to 
the assumed interlayer thickness, which means that the 
IIL is still depleted even for such thin films and no 
diffusion of dopants occurs. The $C$-$V$ curves shown in 
\ref{fig_1}b obey the Mott-Schottky rule as long as no 
significant current appears (between 0 V and -4 V, see Supporting Information). In 
the forward voltage direction thus, a sudden capacitance 
drop appears independently of the IIL thickness at 2.3 V, 
where the geometrical capacitance dominates the $C$-$V$ 
curves since the flat-band condition is reached. The small drop of capacitance 
for a small forward bias is related to the ETL system of cesium doped BPhen. The significant number of trap states
in this material which will be filled under forward voltage conditions causes the drop in capacitance. As one can see e.g. in  \ref{fig_3}b this 
effect does not appear in doped RE68. 
However, in reverse direction, a weak decrease of the capacitance is 
observed as predicted by the Mott-Schottky rule. This 
corresponds to a charge depletion at the HTL/IIL and 
ETL/IIL interfaces. Besides this, a drop of the capacitance 
is obtained as expected from the $I$-$V$ curves for certain 
reverse voltages depending on the IIL thickness caused by a significant injection into the ILL.

\begin{figure}
\includegraphics[scale=0.6]{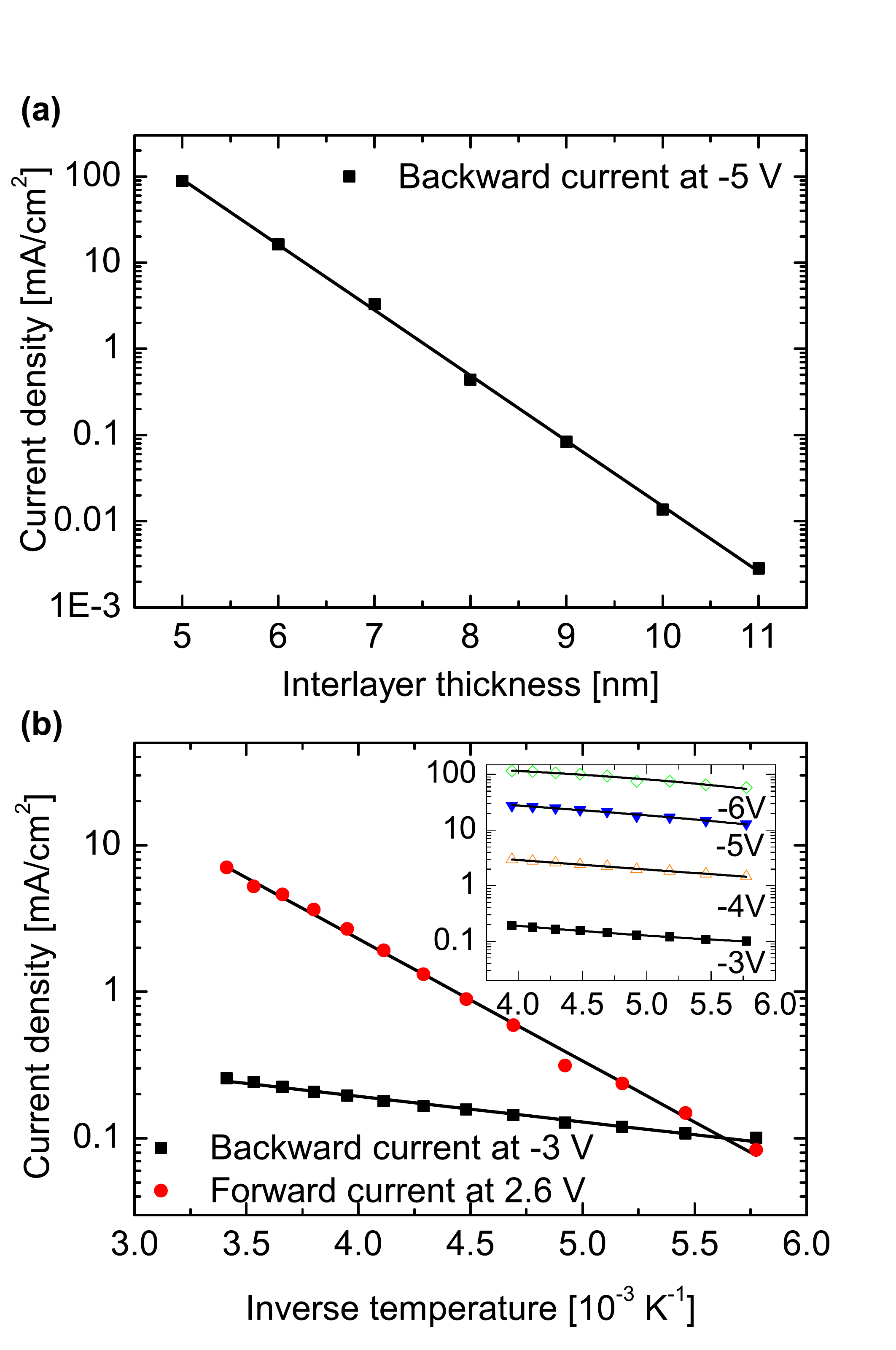}
\caption{ Current density at a reverse voltage 
of -5 V for different interlayer thicknesses. Device structure as in \ref{fig_1} 
(b) Temperature dependence of the current density in 
reverse direction measured at -3 V and in the 
exponential rise of the forward direction at 2.6 V. 
The inset shows the temperature dependence for different backward 
voltages. The devices consist of Al(100 nm)/ 
MeO-TPD:NDP2[4 wt\%](50nm)/ Balq2:NPB[50 wt\%](7nm)/ 
Bphen:Cs(50 nm)/ Al(100 nm) on a cleaned glass substrate.}\label{fig_2}
\end{figure}

To demonstrate the control of the reverse behavior and 
to discuss the possible relevant charge transport mechanisms, we plot in \ref{fig_2} the reverse current for different temperatures 
and IIL thicknesses. For decreasing ILL thickness 
(from 11 nm to 5 nm), a purely exponential current 
dependence is observed (for reverse voltages larger than -3 V), with an increase of the reverse 
current by nearly six orders of magnitude.

A further strong evidence of tunneling is the observed temperature dependence:
In addition to the exponential behavior, we find  a rather weak influence of the temperature 
on the reverse current. In contrast 
to a typical activation energy of the forward current 
($E^{for}_{act}$=(170$\pm$20) meV) for such a $pin$-device, {\cite{Maennig2001}}
 a surprisingly small activation energy of $E^{rev}_{act}$=(30$\pm$10) meV 
is measured in the reverse direction, without any voltage 
dependence (see inset of \ref{fig_2}b). This low reverse 
activation energy can be interpreted as being related to HOMO-LUMO tunneling processes  which  become dominant in this voltage region. 
The difference between this proposed mechanism 
and the effect described by Drechsel et al. {\cite{Drechsel2002}}
can clearly be seen in the weak interlayer thickness dependence of the backward current reported by Drechsel. This is
mainly due to a change in the injection conditions for majority charge carriers at the schottky contact and not related to a tunneling of minority 
charge carriers as reported here.
Concerning the activation energy in reverse direction it should be
mentioned that since we can not neglect the influence of the transport layers to the temperature dependence the
real value of activation energy remains elusive.
However, even if the found temperature and interlayer dependences are a strong evidence for a tunneling mechanism a second effect, the Avalanche breakdown, could also play a role as a possible transport mechanism. 
Since we obtain an exponential current-voltage characteristic for different interlayer thicknesses and for a large voltage region just one kind of these two processes is supposed to be dominant.
Avalanche and Zener breakdown are distinguishable by the field, which needs to be applied to be in the breakdown region. \cite{Sze1966,McAfee1951}  
Thus, for a certain energy gap of the materials used for the diode Zener tunneling should appear for lower fields than Avalanche multiplication. Since we do not obtain a second process in the current-voltage curves for higher voltages the Zener effect is supposed to be the dominant mechanism.

If one assumes a transport mechanism similar to inorganic 
Zener diodes, one can further expect a control of the 
reverse breakdown by the doping concentration of the 
transport layers. \ref{fig_3} shows data a for a diode 
of RE68 designed as a homojunction: the breakdown voltage 
can be shifted to higher values for lower doping concentrations 
as expected. The reason is a change of the 
effective thickness of the intrinsic layer by an additional depletion region at the 
HTL/IIL and the IIL/ETL interface. Although the reverse current has a reduced 
sensitivity to the doping concentration than to the IIL 
thickness, this allows a second independent approach to 
control the breakdown voltage. Furthermore, we also 
obtain no influence on the forward direction of the diode, 
which means that neither the series 
resistance of the transport layers nor the flat-band 
condition have been significantly changed.

$C$-$V$ spectroscopy investigations (see \ref{fig_3}b) once more reveal a Mott-Schottky like behavior for  
different doping concentrations. Obviously, there is an increase of the capacitance for 
a decreasing concentration of dopant molecules. 
Hence, if the thickness of the depletion layer is calculated
from the capacitance, one obtains e.g. 8.3 nm for the 4 wt\% 
of dopants and 11.3 nm for the 0.5 wt\% of dopants 
(including the IIL thickness of 7 nm). This means that 
the depletion region thickness is the sum of the IIL 
and the interface depletion thickness, being sensitive 
to the doping concentration.  
We obtain a similar rule as mentioned before in case of 
the thickness controlled reverse current: the reverse 
breakdown can be shifted by 1 V or the reverse current 
can be increased by one order of magnitude by adding 
one additional nanometer to the thickness of the 
depletion layer (IIL plus interface depletion). 
This doping controlled reverse behavior can also be 
obtained for various concentrations of n-type dopants in ETL 
without any noticeable difference. Instead of NDN1 one can use cesium as 
a freely available n-type dopant (see Supporting Information).

\begin{figure}
\includegraphics[scale=0.6]{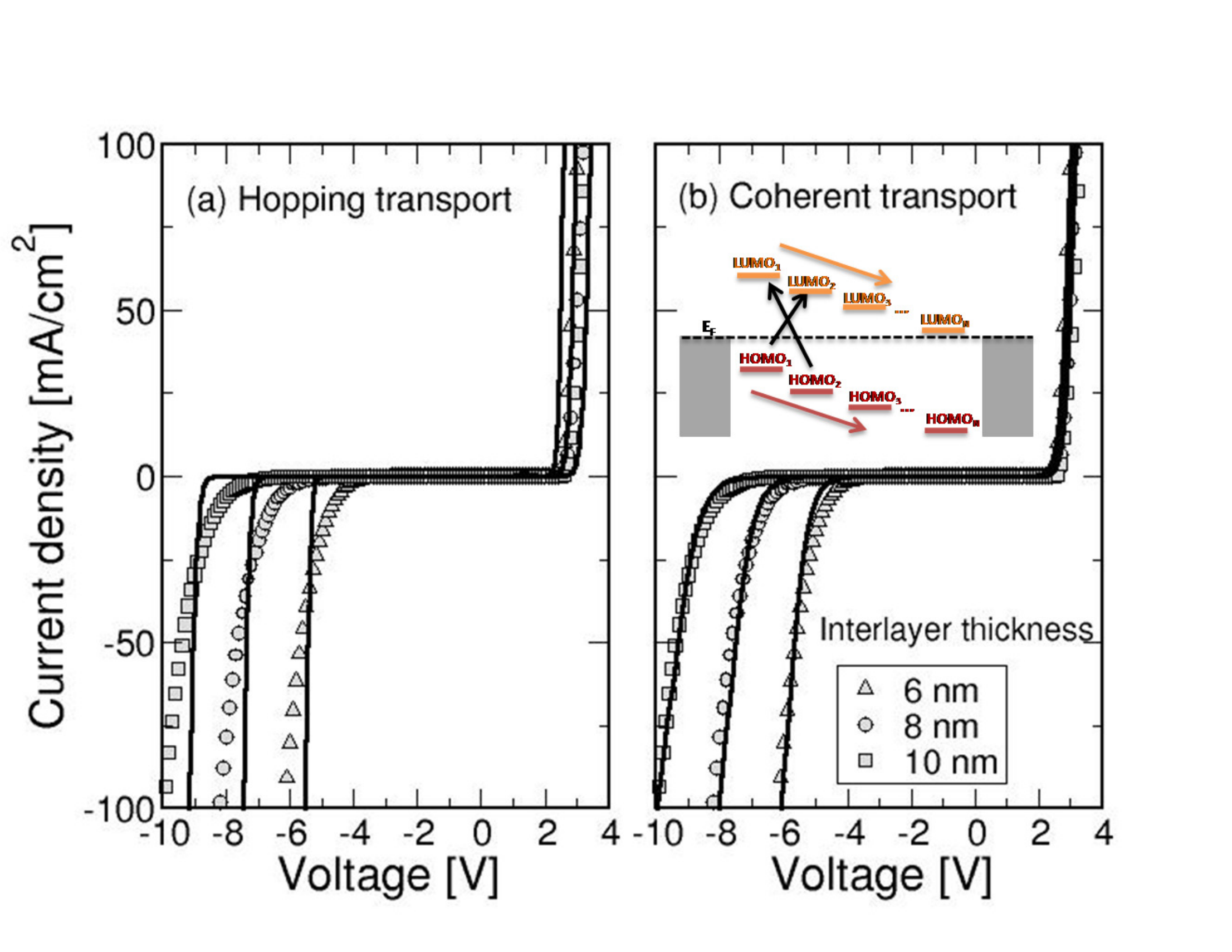}
\caption{Calculated $I$-$V$ curves in the (a) hopping and 
(b) coherent transport regimes compared with the measured characteristics. 
The inset shows a schematic drawing of the two legs electronic ladder used to
 model charge transport in the device. The basic model parameters are: 
(a) $t_0$=150 meV, $t_A^{H-L}$=50 meV, $t_B^{H-L}$=0, $\Gamma_l^L=\Gamma_r^H=\Gamma_l^H=\Gamma_r^L$=25 meV, 
and $\kappa$=0.5, 0.37 and 0.27 eV for $N$=3, 4, and 5; (b) $t_0$=150 meV,
$t_A^{H-L}$=150 meV, $t_B^{H-L}$=50 meV, $\Gamma_l^L=\Gamma_r^H=\Gamma_l^H=\Gamma_r^L$=100 meV, $\beta$=1.2, and 
$\kappa$=0.66, 0.6 and 0.55 eV for $N$=6, 7, and 8. For a definition of the model parameters we also refer the reader to the Supporting Information.} \label{fig_4}
\end{figure}

\textit{Theoretical Modeling.~} The IIL part of the device is not expected to have an ordered crystalline structure, but rather display a high degree of static and dynamic disorder. Since the spatial arrangement of the molecular building blocks in the IIL is not known a priori, a first-principle description of the IIL electronic structure as well as of charge migration~\cite{troisi2010,PhysRevLett.102.116602,citeulike:2842232} in such a system represents a very strong challenge.  Hence, to gain a first qualitative insight into the possible charge transport pathways, we have adopted a minimal model approach. Basically, the model consists of an electronic ladder  with $N$ blocks describing the IIL, each block containing 
two energy levels $\epsilon_{j}^{s}(V)$ whose energetic position is determined not only by the intrinsic electronic structure of a molecular building block but also by the built in field and by the applied voltage (see inset of \ref{fig_4}). The Hamilton operator for this tight-binding model can be written as:
\begin{eqnarray}
 H  &=&  \sum_{s=H,L} \sum_{j}{ \epsilon^s_j (V)}d^+ d_{sj} + \sum_{s=H,L}{t^s}\sum_{j}{( d^+_{sj} d_{sj+1} + h.c.)} \nonumber \\ 
 &+& t^{H-L}_A \sum_{j}{( d^+_{Hj} d_{Lj+1} + h.c.)} + t^{H-L}_B \sum_{j}{(d^+_{Hj} d_{Lj+2} + h.c.)} \nonumber \\
 &+& \sum_{k,\alpha=l,r}{\epsilon_{k\alpha} c^+_{k\alpha} c_{k\alpha} } + \sum_{k,\alpha=l,r}\sum_{j,s=H,L}{t_{k,\alpha,j,s}(c_{k\alpha}^+d_{sj} + h.c.)}.
\label{model}
\end{eqnarray}

The levels are denoted by $s=H$ (highest occupied molecular orbital, HOMO) 
and $L$ (lowest-unoccupied molecular orbital, LUMO) and are thus 
representative of the frontier orbitals of a given molecular
unit. Their energetic position is assumed to be given by the  linear relation:
$\epsilon^{s}_{j}(V)=\epsilon^{s}-\kappa j +\frac{j}{N}eV$,
where $\epsilon^s$ is the bare onsite energy of the molecular orbital (assumed to be site-
independent for simplicity). The local H-L gap
$\Delta_{H-L}=\epsilon^{H}-\epsilon^{L}\sim$ 3.0eV is taken from the experimental
energy diagram profiles. The built-in electric field is modeled as a linear ramp with
slope $-\kappa$, which is supported by the current  experiments as well as by photo-electron spectroscopy measurements in 
organic $pin$-diodes.~\cite{Olthof2009} The second, third, and fourth
summands in Eq. \ref{model} describe different electronic couplings within the ladder: intra-
strand nearest-neighbor couplings (second term) as well as nearest- and next-nearest
neighbor inter-strand interactions (third and fourth terms). If not stated otherwise, we
will assume for the sake of simplicity in the calculations $t^H = t^L =t_ 0$ and $t_{A}^{H-L}$ and $t_{B}^{H-L}
=\eta t_0$, with $0\le\eta<1$. All electronic parameters can in principle be obtained from
microscopic first-principle calculations; this goes however beyond the scope of the
present minimal model approach. The last row in Eq. \ref{model} describes the electronic structure
of the $\alpha=l$ (left) and $r$ (right) electrodes and the coupling between them and the ladder.
The specific electronic properties of the electrodes will not appear explicitly in our
treatment; they are considered only as sinks of charges and will be included in the calculations via phenomenological parameters. 
The electrode-ladder
interaction, given by the $t_{k\alpha,sj}$ matrix elements is in general not  limited to the ladder
sites nearest to the electrodes but it can have a longer range. This will qualitatively account for the fact that the real structural molecular arrangement in the active region does not in general have a linear chain topology $-$it is expected to  rather be a complex network$-$ so that
interactions do not necessarily have to be confined to the electrodes' nearest neighboring molecular blocks.

The \textit{Ansatz} for  $\epsilon_{j}^s(V)$ allows 
to account qualitatively for two different physical scenarios: 
with increasing $V>0$, the built-in field $\kappa$ can be 
progressively counter-acted by the voltage-dependent term $(j/N)eV$, the energy levels become  increasingly aligned and the current can dramatically 
grow due to the reduction of tunnel barriers between  the sites. For negative bias, transport is blocked up 
to a critical bias $-V^{*}$; the energy scale $|eV^{*}|$ is  roughly related to the situation where the $H$-level 
at a site $j$ and the $L$-level at site $j+1$ become nearly 
resonant (additionally, at least the first site of the 
$H$-strand should move above the left chemical potential 
in order to have a current flow). Once this happens,  an exponential current onset can occur. We have used 
two different modeling strategies based on two possible  different microscopic transport mechanisms: hopping 
transport (HT) and coherent tunneling (CT). HT seems 
to be reasonable, since it is expected that the active 
device region displays some degree of static and dynamic 
disorder which may suppress coherent tunneling pathways 
over long distances. We have used a reduced density 
matrix approach~\cite{RD_book} (for HT) and the Landauer formalism~\cite{datta_book} 
to compute the $I$-$V$ characteristics of the model in the 
CT regime (see the Supporting Information for details of the calculation).

 The results (scaled by an arbitrary factor) 
are displayed in \ref{fig_4} in comparison with the experiments. 
It is obvious that the weak (strong) thickness dependence 
for positive (negative) voltages can be qualitatively 
described within both transport regimes. Notice, however, 
the different number of sites needed in both regimes to 
account for the experimental curves, This is related to 
the fact that tunneling processes can still be  
efficient for short lengths, so that almost no 
zero-current gap is obtained. Only with increasing length tunneling transport is blocked at low bias.

Some basic features of the transport process can be summarized as follows: for positive voltages the onset of current takes
place approximately at the critical bias where the first site of the HOMO-strand moves
above the left chemical potential $\mu_{l}$ (which is kept fixed at $\mu_l=E_F$ while $\mu_r=E_{\textrm{F}}+eV$).
Under this condition, charge transport can take place. A further increase of the bias
leads to an increase of the current till all the electronic levels along the ladder become
aligned (the external potential cancels the built-in field $\kappa$). This provides the maximum
current in this model; for larger positive voltages the misalignment of the levels
develops again and negative differential resistance can eventually be obtained.
For negative voltages, the current is almost zero in the case of coherent tunneling, since this process
is very sensitive to the presence of energy gaps along the ladder. 
This behavior can be lifted by allowing for long-range interactions between the electrodes and  the ladder states (see Eq.~7 in the Supporting Information). On the contrary,
hopping transport can also take place in the negative bias regime with only nearest-neighbor interactions. 

Our results suggest that the effective microscopic transport mechanism 
may include both HT and CT contributions, each of them 
becoming dominant in different voltage regions. This may 
be the reason for the different slopes in the $I(V=const.,T)$ 
vs. $T^{-1}$ plots for positive and negative voltages (\ref{fig_2}). 
A pure CT model cannot account for a $T$-dependence; 
HT contributions can however explain it. In the HT 
regime, a different qualitative behavior for positive and negative voltages is expected on 
the basis of a simple argument: 
the nearest neighbor inter-site hopping rates within the ladder include (see Eq.~4 in the Supporting Information)
thermal factors $~exp(-(\epsilon_{j+1}^s-\epsilon_j^{p})/k_BT)$, whenever 
$\epsilon_{j+1}^s-\epsilon_j^{p}>0$ $(s,p=H,L)$. For $s\ne p$ (sites on different strands), 
one obtains $\epsilon_{j+1}^s-\epsilon_j^{p}=\Delta_{H-L}-\kappa+eV/N$; from here it is 
obvious that for $V>0$ the temperature dependence is mainly 
determined by the gap $\Delta_{H-L}\ll k_{B}T$ (large slope). For 
negative bias, $\Delta_{H-L}$ can be partially compensated by the 
term $-\kappa-e|V|/N$ and thus the temperature dependence may 
be weakened. A more accurate quantitative analysis 
requires a systematic experimental study of the 
temperature dependence, taking also the influence 
of the transport layers into account.

\textit{Conclusions.~} In this Letter, we unambiguously observe signatures of Zener tunneling in a organic molecular semiconductor:
The reverse currents in organic $pin$-diodes show geometry, field and temperature 
dependence in excellent agreement with this tunneling effect not observed so far in 
such systems.  The realized $pin$-Zener diodes have excellent device properties and allow 
important application possibilities, e.g. to reduce  crosstalk in passive matrix arrays. 
The basic effect of valence- to conduction-level tunneling  could be used in further device concepts, such as devices 
employing tunable resonant tunneling. Using a minimal model Hamiltonian, we can describe semi-quantitatively the measured electrical  response. However, both limiting cases investigated here, incoherent hopping  or coherent tunneling, can describe the observed experimental features. It therefore seems appealing to expect that the dominant charge transport mechanism may sensitively depend on the sign of the applied bias. A further point to be considered is the coupling to dynamical degrees of freedom which may induce polaron formation, so that  the transport process will be rather mediated by dressed electrons or holes. We have not taken these issues in our model which is only addressing  electronic degrees of freedom. Further work in this direction including  detailed first-principle electronic structure calculations is thus needed.

\acknowledgement  The authors thank Novaled AG, the free state of Saxony 
and EU (EFRE) for financial support under project 
NKOE (12712) and Novaled for providing the HTL dopant NDP2 
and the ETL dopant NDN1.
This work was also supported by the Volkswagen Foundation 
through grant no. I/78-340, by the cluster of excellence 
of the Free State of Saxony European Center for Emerging 
Materials and Processes Dresden ECEMP/A2, and by WCU 
(World Class University) program through the Korea 
Science and Engineering Foundation funded by the 
Ministry of Education, Science and Technology 
(Project No. R31-2008-000-10100-0).

\bibliography{lit_zener}

\suppinfo
{The Supporting Information Section includes preparation methods, additional $I$-$V$, $C$-$V$, $C$-$f$ and Mott-Schottky-curves for 
the free available materials. Furthermore we present a more detailed description of the
microscopic model applied to describe the charge transport characteristics in the organic Zener diode. }

\end{document}



\section{Supporting Information}
\subsection{Experimental Methods}

\textit{Preparation.~} The devices are prepared on cleaned (supersonic treatment in acetone, ethanol and
isopropanol for 5 minutes and O2 plasma etching) glass substrate by thermal
evaporation of metals and organic materials under high-vacuum conditions (base
pressure  $< 10^{-5}$ Pa) without breaking the vacuum. The layer evaporation is done in the
following sequence: Aluminium (bottom contact, thickness 50 nm), HTL (50 nm), IIL,
ETL (50 nm), Aluminium (top contact 50 nm). The thicknesses are confirmed by quartz
crystal monitoring and the active area is 6.38 mm$^2$.

\begin{figure}
\includegraphics[scale=0.4]{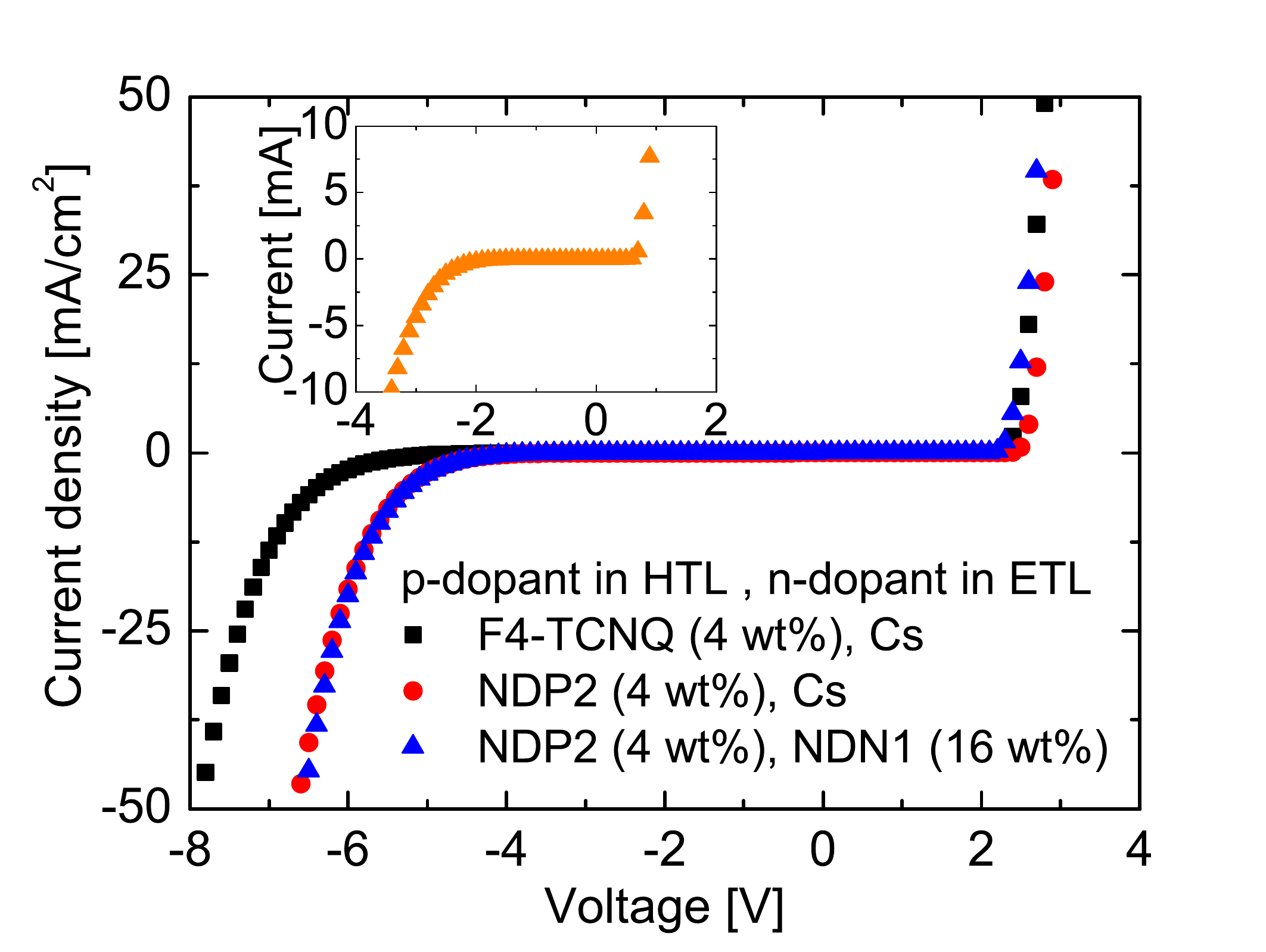}

\caption{ I-V curves for different p-type and n-type dopants. The device
consists of an intrinsic interlayer (7nm) of Balq2:NPB sandwiched between a
hole and an electron transport layer (both 50nm) of MeO-TPD and Bphen,
respectively. The inset shows the I-V curve of a commercial inorganic silicon
Zener diode with a breakdown voltage of 2.7V.
}
\label{fig1}
\end{figure}

The main materials used are: MeO-TPD: N,N,N9,N9 tetrakis (4-methoxyphenyl)- benzidine. 
NPB: N,N9- di(naphthalen-1-yl)- N,N9)-diphenyl- benzidine. 
BAlq2: aluminium- (III)bis (2-methyl-8-quninolinato)- phenylphenolate. 
Bphen: 4,7-diphenyl-1,10-phenanthroline.
RE68: Tris (1-phenylisoquinoline) iridium(III). For the 
p-type doping, of the HTL the material NDP2 (Novaled AG, Dresden) 
can be used as well as the material F4TCNQ: 2,3,5,6- tetrafluoro-7,7,8,8- tetracyanoquino- dimethane 
as freely available p-dopant giving equivalent 
electrical properties (see \ref{fig1}). The n-type 
doping of the ETL can be achieved by co-evaporation of 
Bphen and pure Caesium. Likewise, the material NDN1 
(Novaled AG, Dresden) acts as an effective n-type dopant 
(see  \ref{fig1}).

For the electronic characterization a Keithley 2400 SMU is used for $I$-$V$
investigations while a HP 4284 $LCR$-meter is used for the $C$-$V$ and $C$-$f$ studies. The $I$-$V$
curves are done with a voltage sweep (from -10V to 4V and back) and the $C$-$V$
measurements are prepared at 1 kHz with an alternating sine signal (RMS value) of 20 mV.

\textit{Organic Zener diode with different molecular dopants.~} Owing to the highly efficient
doping of the hole and electron transport layers, the organic Zener diode becomes a
reproducible and stable device with a low series resistance. Also, doping provides a
second independent way to control the reverse behaviour.
We can show (see Figure 1), that the effect of the reverse breakdown appears
independently of the used dopant. However, due to the different efficiencies of the
dopant materials, one can observe an influence of the used dopants on the reverse
direction. The number of charged dopant states at the interface between the doped
layers and the intrinsic interlayer determines the thickness of the depletion region which
controls the reverse current. Accordingly, one can further extract from Figure 1 that the
molecular doping of MeO-TPD by F4TCNQ is less efficient than by NDP2 (for the
same mass-ratio,corresponding to a molar percentage of approximately 8mol\% for F4TCNQ 
and 6mol\% for NDP2, respectively). Nevertheless, the series resistance of the
transport layers is not significantly affected and hence no change in the forward
direction is obtained.

\textit{C-f and C-V investigations.~}  In the frequency range from 50Hz to 1MHz, the organic
Zener diode provides a phase and impedance response (see Figure 2 a) which fits to a
simple RC-unit as the equivalent circuit. Since the doped transport layers have a small
series resistance and also a small geometrical capacitance (thickness 50nm) the RC-unit
belongs to the intrinsic interlayer, which has the dominant RC-time in this stack (inset
of Figure 2 a). One can thus identify the capacitance with the geometrical thickness of
the depleted area. The series resistance describes Ohmic losses along the contacts and
the transport layers, while the resistance parallel to the capacitor models the influence of
charge recombination and leakage currents for the reverse biased junction.
This correlation between the capacitance and the thickness of the depletion, which is the
sum of the interlayer thickness and the interface depletion, can be seen in Figure 2 b.
\begin{figure}
\includegraphics[scale=0.4]{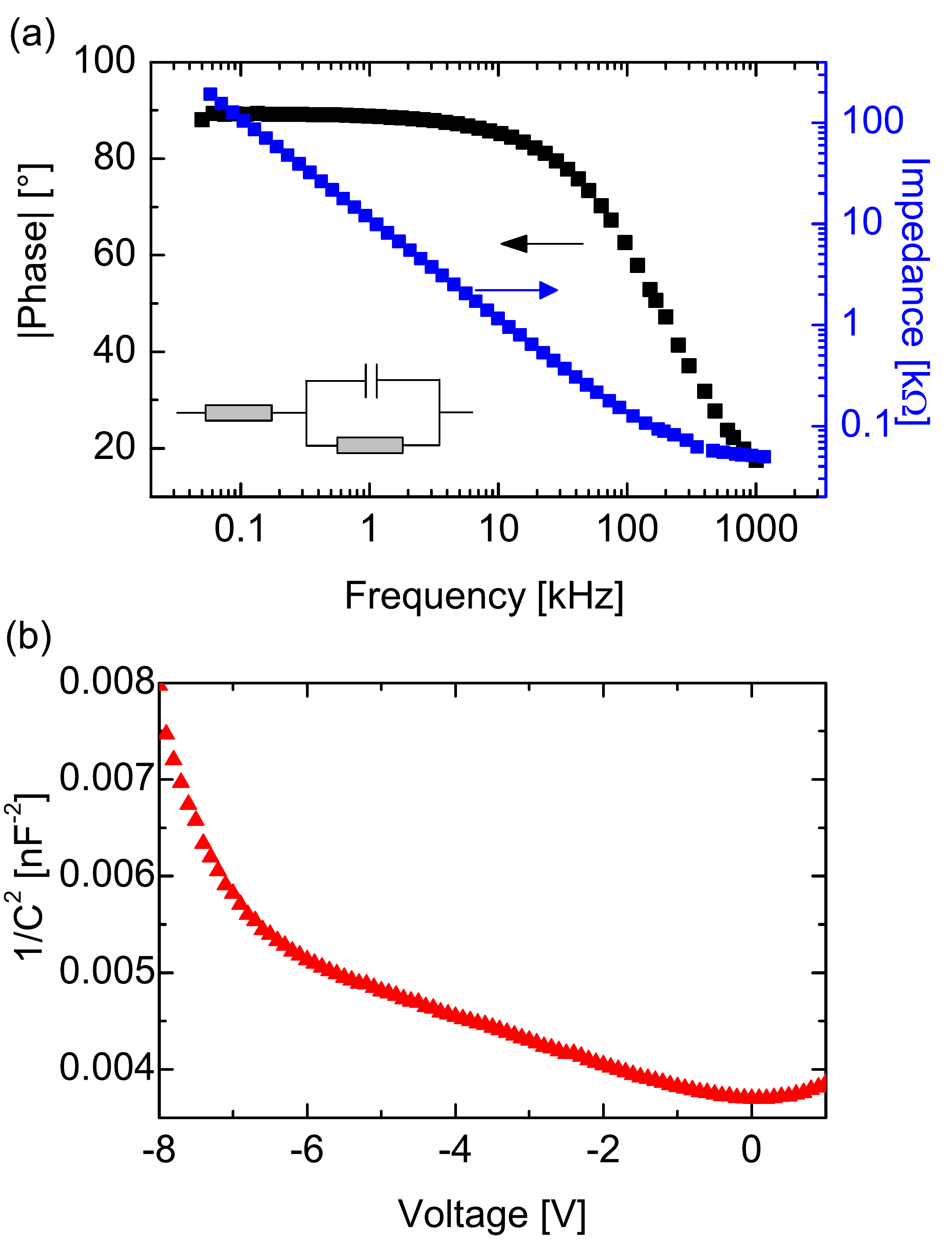}
\label{fig2}
\caption{ C-f and C-V curves of the organic Zener diode. (a) C-f
measurements of the organic Zener diode. In accordance to the frequency
dependence of the phase and the impedance one can assume a simple RC-unit
as the equivalent circuit (inset). (b) Mott-Schottky plot of the device shown in
(a). The device consists of Al(100nm)/ MeO-TPD:F4TCNQ[4wt\%](50nm)/
Balq2:NPB[50wt\%](7nm)/ Bphen:Cs(50nm)/ Al(100 nm)
} 
\end{figure}
For a reverse voltage between 0V and -6V we obtain a linear dependence in the Mott-
Schottky plot. This means, the C-V curves follow
\begin{equation}
1/C^2 = \frac{2}{e \epsilon \epsilon_0 A} \frac{N_A + N_D }{N_D N_A}(V-V_{bi})+d_i ^2 .  
\end{equation}
Here, N$_A$ and N$_D$ are the densities of charged dopant states, d$_i$ is the IIL thickness, A the
active device area, V the applied voltage, and V$_{bi}$ the built-in voltage of the diode.
Accordingly, there is a voltage dependent depletion at the interfaces between the
transport layers and the intrinsic interlayer as predicted by the Mott-Schottky rule. This
means that the assumption of two rectangular space charge regions separated by an
intrinsic layer is fulfilled in our pin-structure. With respect to this, we can expect a
linear voltage drop over the intrinsic layer.
It should be further mentioned that the Mott-Schottky rule is not valid for the forward
biased device and also in the range of high negative voltages. The contributions of drift,
diffusion current and recombination processes, which are not taken into account in this
simple model, cause a collapse of the assumed equivalent circuit.

\subsection{Theoretical Methods}
Using the model introduced in the main text, Eq.~1, we can formally treat different transport scenarios. In the following  we will address two limiting cases: incoherent, thermally assisted  transport and coherent tunneling. 

\textit{Incoherent transport.~} The model in Eq.~1 has been applied to calculate the I-V
characteristics for different lengths N. We may expect that static and dynamical disorder
in the active region will make a fully coherent transport scenario very unlikely. Only for
narrow regions (short length scales) quasi-coherent transport might still be efficient, though this may also
depend on the voltage region. The electrical response of the ladder will thus be first
studied in the limit of complete coherence loss. For this, a reduced density matrix
approach~\cite{RD_book}  has been used and only its diagonal components (site populations) will be considered. 
Off-diagonal terms are assumed to have decayed on any relevant time
scale. As a result, a set of classical rate equations can be derived (setting in this case the
next nearest neighbor coupling $t_B^{H-L}$ to zero, but keeping $t_A^{H-L}\neq 0$):
\begin{eqnarray}
\frac{d}{dt}P_{H,n}(t) & = &\Gamma_{1,n}\left(f_1(\epsilon_n^H)-P_{H,n}(t)\right) + \Gamma_{2,n}\left(f_2(\epsilon_n^H)-P_{H,n}(t)\right) +\gamma_{HH,n+1}P_{H,n+1}(t)+\gamma_{HH,n-1}P_{H,n-1}(t) \nonumber \\  
 &+&\gamma_{HL,n-1}P_{L,n-1}(t) +\gamma_{HL,n+1}P_{L,n+11}(t) - \left(\gamma_{HH,n-1}+\gamma_{HH,n+1} +\gamma_{HL,n-1}+ \gamma_{HL,n+1} \right) P_{H,n}(t) \nonumber \\
\frac{d}{dt}P_{L,n}(t)  &=& \Gamma_{1,n}\left(f_1(\epsilon_n^L)-P_{L,n}(t)\right) + \Gamma_{2,n}\left(f_2(\epsilon_n^L)-P_{L,n}(t)\right) +\gamma_{LL,n+1}P_{L,n+1}(t)+\gamma_{LL,n-1}P_{L,n-1}(t) \nonumber \\ 
 &+&\gamma_{LH,n-1}P_{H,n-1}(t) +\gamma_{LH,n+1}P_{H,n+11}(t) - \left(\gamma_{LL,n-1}+\gamma_{LL,n+1} +\gamma_{LH,n-1}+ \gamma_{LH,n+1} \right) P_{L,n}(t). \nonumber \\
\end{eqnarray}

Here, $P_{H,n}(t)$ and $P_{L,n}(t)$ are the corresponding populations and the transition rates $\gamma$ can
be expressed in terms of the tight-binding hopping parameters along the ladder of the model Hamiltonian in Eq.~1 (main text). $f_{s=1,2}(\epsilon^{H,L}_n)$ are
equilibrium Fermi functions of the electrodes evaluated at energies $\epsilon^{H,L}_{n}$. The
parameters $\Gamma_{sn}=\Gamma_{s,n=m}$ describe phenomenologically the charge hopping rates into and out of the electrodes
and can be formally expressed via the coupling terms $t_{k\alpha,sj}$ using Fermi's Golden Rule.
Assuming them to be energy-independent constants within the so called  wide-band limit,~\cite{datta_book} we make the following $Ansatz$:
\begin{eqnarray}
\Gamma_{l,nm}&=&\delta_{n1} \delta_{m1}\Gamma_{l}^H, \hspace{1.7cm} \Gamma_{r,nm}=\delta_{n=N} \delta_{m=N}\Gamma_r^H \hspace{0.8cm} \text{if} \; \; n\in H \; \text{strand} \nonumber \\
\Gamma_{l,nm}&=&\delta_{n=N+1} \delta_{m=N+1}\Gamma_{l}^L, \; \; \; \;\Gamma_{r,nm}=\delta_{n=2N} \delta_{m=2N}\Gamma_r^L \; \; \; \; \; \; \text{if} \; \; n\in L \; \text{strand}.
\end{eqnarray}
For the hopping rates we further assume:~\cite{PhysRevB.81.115203}
\begin{eqnarray}
\gamma_{sp,j}&=&(t^{sp})^2 e^{-\frac{\epsilon^s_p(V)-\epsilon_m^p(V)}{k_B T} } \; \; \text{if} \; \; \epsilon^s_p(V)-\epsilon_m^p(V)>0, \nonumber \\
\gamma_{sp,j}&=&(t^{sp})^2  \hspace{2.2cm}  \text{if} \; \; \epsilon^s_p(V)-\epsilon_m^p(V)<0, 
\end{eqnarray}
where $s$ and $p$ denote the $H$ or $L$ strands, while $n,m$ are site indexes along the strands or
between them. Depending on the energy difference between the sites there may be an
energetic barrier to climb or an activation-less hopping process. In the stationary
regime, $dP_{H,n}/dt= dP_{L,n}/dt =0$, and a linear set of equations for the populations can be
obtained. Finally, using the definition of the stationary electrical current coming from
lead $\alpha$, $I_\alpha (V)=-e<dN_\alpha/dt>$, one finally gets
\begin{eqnarray}
I_{tot}(V)&=& \frac{e}{\hbar}\left( \Gamma^H_r\left[f_r(\epsilon_N^H,V)-P_{H,N}(V)\right] + \Gamma^L_r\left[f_r(\epsilon_N^L,V)-P_{L,N}(V)\right]\right) \nonumber\\
&-&\frac{e}{\hbar}\left( \Gamma^H_l\left[f_l(\epsilon_1^H,V)-P_{1,H}(V)\right] + \Gamma^L_l\left[f_l(\epsilon_1^L,V)-P_{1,L}(V)\right]\right). 
\end{eqnarray}
\textit{Coherent tunneling.~}To explore to which degree coherent charge propagation could
contribute to the observed length dependence of the $I$-$V$ characteristics, we have also used
the Landauer theory~\cite{datta_book} for the model of Eq.~1. Hereby, the key quantity to be computed is the energy dependent quantum mechanical transmission probability $T(E)$, which encodes the electronic structure of the system as well as the coupling to the electrodes. $T(E)$ can be computed 
according to the standard expression $T(E)=Tr[G^{a}(E)\Gamma_{r}G^{r}(E)\Gamma_{l}]$. $G^{r,a}(E)$ denote
retarded (advanced) ladder Green functions including the interaction with the electrodes.
The electrode-system coupling matrices $\Gamma_{l,r}$ are defined below in \ref{gamma}. The Green
functions can be calculated via Dyson's equation.~\cite{Ryndyk2009} 
The electrical current through the
system is then obtained  (with $\mu_{l} =E_F$ and $\mu_{r}=E_F+eV$ being the chemical potentials of the
electrodes) as:
\begin{equation}
I(V)=\frac{2e}{h}\int dE\left[ f(E-\mu_{l})-f(E-\mu_{r})\right]T(E).
\end{equation}
However, in order to have a non-zero current for negative voltages, it turned out that
coupling beyond nearest neighbors, i.e. $t_B^{H-L}\neq 0$, as well as a long-range coupling $\Gamma_{l,r}$ to the
electrodes was required. The latter was now modeled as:
\begin{eqnarray}
\Gamma_{l,nm}&=&\delta_{nm} \Gamma_l^H e^{\beta n}, \; \Gamma_{r,nm}=\delta_{nm} \Gamma_r^H e^{\beta(N-n)} \; \; \text{if} \;  \; n\in H \; \text{strand} \nonumber \\
\Gamma_{l,nm}&=&\delta_{nm} \Gamma_l^L e^{\beta n}, \; \Gamma_{r,nm}=\delta_{nm} \Gamma_r^L e^{\beta(N-n)} \; \; \text{if} \; \;  n\in L \; \text{strand}.
\label{gamma}
\end{eqnarray}
The reason for the strong current suppression at negative biases is related to the fact that in a Stark-ladder (linear slope in the onsite energy profile along the chain) as that used in the present
calculations all electronic states are localized in the thermodynamic limit, so that for
negative bias (where the slope of the ladder is very large) charge tunneling is essentially blocked. The calculated $I$-$V$ curves
are shown in the main text, see Fig. 4. The theoretical curves were scaled by an arbitrary
factor, since the experiments measure current densities and in our approach only
currents are calculated. Hence, a direct quantitative comparison between theory and
experiments turns out to be very difficult in general. Nevertheless, the trends in the
length dependence can also be approximately described by using this coherent transport model.

In general, it may also be possible that
both coherent and incoherent pathways are competing on different bias regimes. To
clear this point requires however a more detailed microscopic study which should
include a first-principle estimation of the different parameters used in the model as well
as the inclusion of a realistic arrangement of the molecular units in the device region.

\bibliography{Si_lit}